\documentclass[a4paper,11pt]{article}
\usepackage{pos}

\title{An enhanced track reconstruction for the CMS experiment at the HL-LHC}

\author*[a]{Mario Masciovecchio}
\onbehalf{on behalf of the CMS Collaboration}

\affiliation[a]{University of California, San Diego,\\
9500 Gilman Dr., La Jolla, CA 92093, United States of America}

\emailAdd{mario.masciovecchio@cern.ch}

\abstract{This contribution presents the new phase-2 CMS baseline tracking strategy currently proposed for online event reconstruction, which can be similarly useful in offline reconstruction under the extreme pileup conditions of the High-Luminosity LHC (HL-LHC). By unifying diverse algorithmic paradigms and integrating machine learning (ML) techniques into a coherent sequence, CMS can maintain high-efficiency and high-resolution tracking while improving the fake track rate and the computational performance, a crucial aspect especially for the HL-LHC. The approach combines GPU-optimized algorithms (Patatrack, LST) with vectorized CPU methods (mkFit), exploiting modern hardware to maximize throughput. This heterogeneous strategy reduces the resource requirements and enhances the physics reach by incorporating a global displaced track reconstruction and extending sensitivity to rare, interesting signatures, such as long-lived particles.}

\FullConference{43rd International Conference on High Energy Physics (ICHEP 2026)\\
30 July  to 5 August , 2026\\
Natal, Brazil\\}

\begin{document}
\maketitle

\section{Introduction}

The High-Luminosity LHC (HL-LHC) will open an era of new
opportunities for discovery and for precision measurements.
Reaching its physics goals requires CMS to operate with up to 200 simultaneous proton-proton interactions 
per bunch crossing (pileup, PU), against an average of about 60 in Run~3.
Track reconstruction is one of the steps of event reconstruction that are most directly affected, 
because its combinatorics grows superlinearly with PU.
With $\langle\mathrm{PU}\rangle = 200$, a single event will contain $\mathcal{O}(10^5)$ tracker hits and
$\mathcal{O}(10^4)$ charged-particle trajectories.
Since essentially every downstream physics object relies on tracks,
any degradation in tracking performance propagates to the entire physics program.
With finite computing resources,
leaving the algorithms unchanged would mean failing to keep up with the
data collection rate.

The CMS response combines several complementary strategies: exploiting the
granularity of the upgraded tracker and multi-core CPUs through parallelized
and vectorized algorithms, complementing them with GPUs through heterogeneous
computing models and GPU-friendly algorithms, and extending the use of machine
learning (ML).
The developments reported here were driven by the needs of the
High-Level Trigger (HLT), where the timing constraints are tightest, but the
resulting sequence largely overlaps with and is being exported to offline
reconstruction.
Performance is quantified by the \emph{efficiency}, the fraction of simulated tracks from the hard-scattering
vertex associated with at least one reconstructed track, the \emph{fake rate},
the fraction of reconstructed tracks that are not associated to any simulated track, 
and the \emph{duplicate rate}, the fraction of reconstructed tracks associated more than once with the
same simulated track;
a reconstructed track is associated with a simulated particle when more than 75\% of its hits originate from it.
All results below use simulated $\mathrm{t\bar{t}}$ events at $\sqrt{s}=14$~TeV with $\langle\mathrm{PU}\rangle = 200$.

\section{The CMS tracker at the HL-LHC}

A completely new tracker will be installed at the center of CMS for the
HL-LHC~\cite{tktdr}. The Inner Tracker (IT) is built entirely from pixel
sensors and comprises 4 barrel layers and $12\times2$ endcap disks; the Outer
Tracker (OT) comprises 6 barrel layers and $5\times2$ endcap disks, built from
modules pairing a strip sensor with either a macro-pixel (``PS'' modules) or a
second strip sensor (``2S'' modules).
Extending pixel sensors out to $r \simeq 60$~cm significantly improves the handling of combinatorics.
Each OT module consists of two closely spaced sensors read out together:
their hits can be used to form a \emph{stub}, or
mini-doublet (MD), whose opening angle gives a local estimate of the track
transverse momentum ($p_{\mathrm{T}}$). These ``$p_{\mathrm{T}}$ modules''
suppress low-$p_{\mathrm{T}}$ combinatorics and enable tracking already at the
Level-1 trigger.


\section{A single-iteration heterogeneous sequence}

Track reconstruction traditionally proceeds in three steps: \emph{seeding},
where seeds are formed from hits on a subset of tracker layers;
\emph{building}, where seeds are extended using the full tracker information;
and \emph{fitting}, from which the track parameters are extracted, optionally
followed by a quality-based selection.
The \emph{legacy} reconstruction runs this sequence over several iterations, 
masking the hits used by the tracks selected in iteration $N$ before iteration $N+1$; 
for the HL-LHC, the legacy HLT configuration consisted of two iterations, seeded respectively by IT-only
quadruplets and IT-only triplets.
The HL-LHC HLT baseline has recently been updated to a \emph{single-iteration} approach~\cite{dp026}
relying on three algorithms: Patatrack and Line Segment Tracking (LST) for seeding, and mkFit
for building (and fitting). The same strategy is being exported to offline tracking,
where mkFit is already used for building in the first legacy tracking iteration.

\subsection{Patatrack}

The Patatrack~\cite{patatrack} algorithm is a parallel, hardware-agnostic pixel track
reconstruction algorithm used by CMS since Run~3 for seeding at the HLT~\cite{dp2022hlt},
offloaded to GPUs on a heterogeneous farm. After the pixel local
reconstruction, charge clusters on adjacent layers are linked into $n$-tuplets
by a Cellular Automaton. These $n$-tuplets are then cleaned, fitted, 
and selected according to quality criteria.
For the HL-LHC, the algorithm has been extended to use the
pixel sensors of the OT in addition to the IT~\cite{dp2025pixel,dp114}, returning
fitted $n$-tuplets with at least four hits in total and at least two in the IT.

\subsection{Line Segment Tracking}

The LST~\cite{dp029} is a parallel, hardware-agnostic algorithm designed around the
Phase-2 OT geometry, whose guiding principle is to link short, highly localized
objects into progressively longer ones; because each link involves only nearby
detector elements, the algorithm is naturally parallel and maps well onto GPUs.
A line segment (LS) is a linked pair of MDs in neighboring layers,
a triplet (T3) a linked pair of LSs sharing an MD, a quadruplet (T4) a linked pair of T3s sharing an LS, 
and a quintuplet (T5) a linked pair of T3s sharing an MD.
Pixel tracks from Patatrack (IT-only or OT-extended) enter as pixel line segments (pLS) 
and are combined with OT objects into pT3 and pT5 candidates, the output being a collection of pT5s,
pT3s, T5s, T4s and unassociated pLSs. 
Fakes and duplicates are suppressed by connection maps, geometric criteria and dedicated ML
selections~\cite{dp2025lstml}. Crucially, because LST builds OT-only
candidates, it provides displaced track reconstruction globally rather than
only in dedicated iterations.

\subsection{mkFit}

The mkFit~\cite{mkfit} algorithm is a vectorized and parallelized implementation of
Kalman-filter track finding and fitting for multi-core CPUs. Parallelization is
handled by Intel TBB; vectorization is obtained through SIMD directives in the
propagation and, for the Kalman operations, through the \emph{Matriplex}
library, a code generator for optimized small-matrix operations. This is enabled by a
generalized description of the detector geometry and its traversal, so that
sets of track candidates undergo the same operations in lockstep. 
In Run~3, mkFit was used to built more than 90\% of all reconstructed offline
tracks with $p_{\mathrm{T}} > 0.5$~GeV: physics performance was retained or
improved compared to legacy tracking, event throughput increased by 10--15\%, and the total track
reconstruction time fell by about 25\%~\cite{dp2022mkfit}. Since 2025, it was
also used at the HLT for track building.

\section{Physics and computing performance}

\begin{figure}[htbp]
  \centering
  \includegraphics[width=0.32\textwidth]{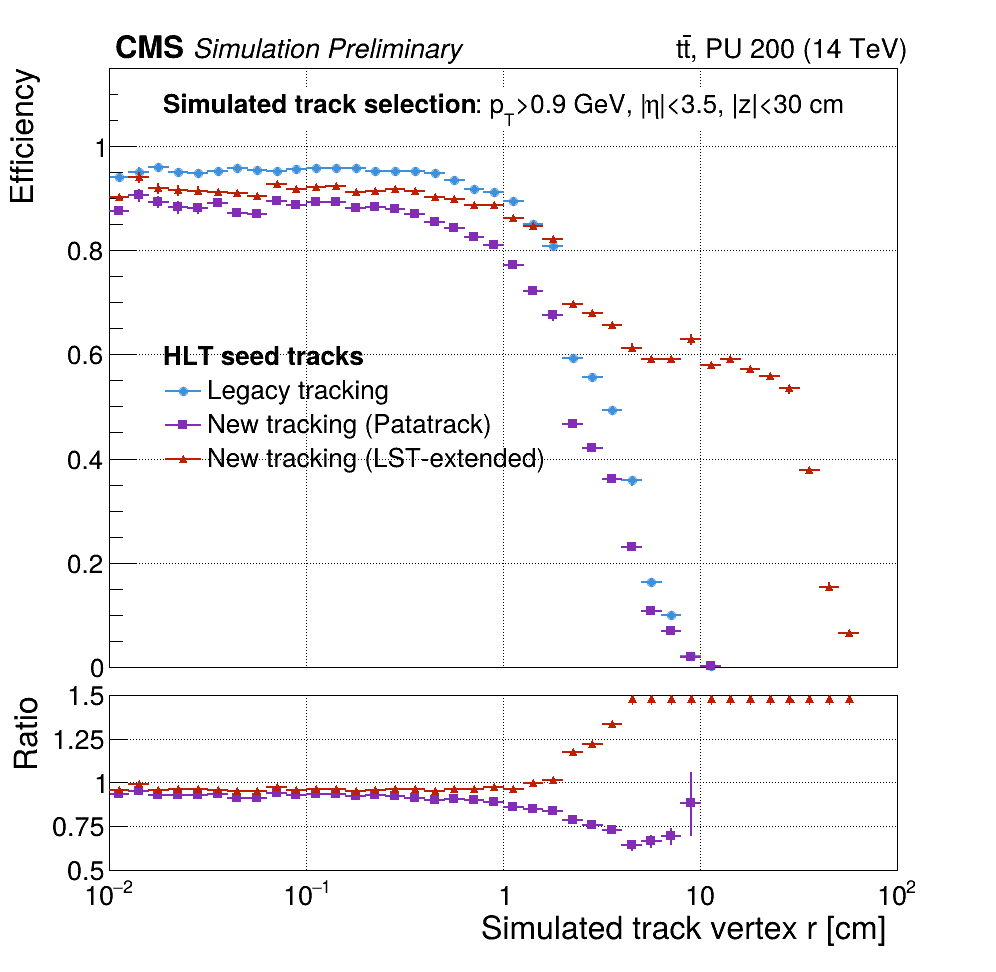}
  \includegraphics[width=0.32\textwidth]{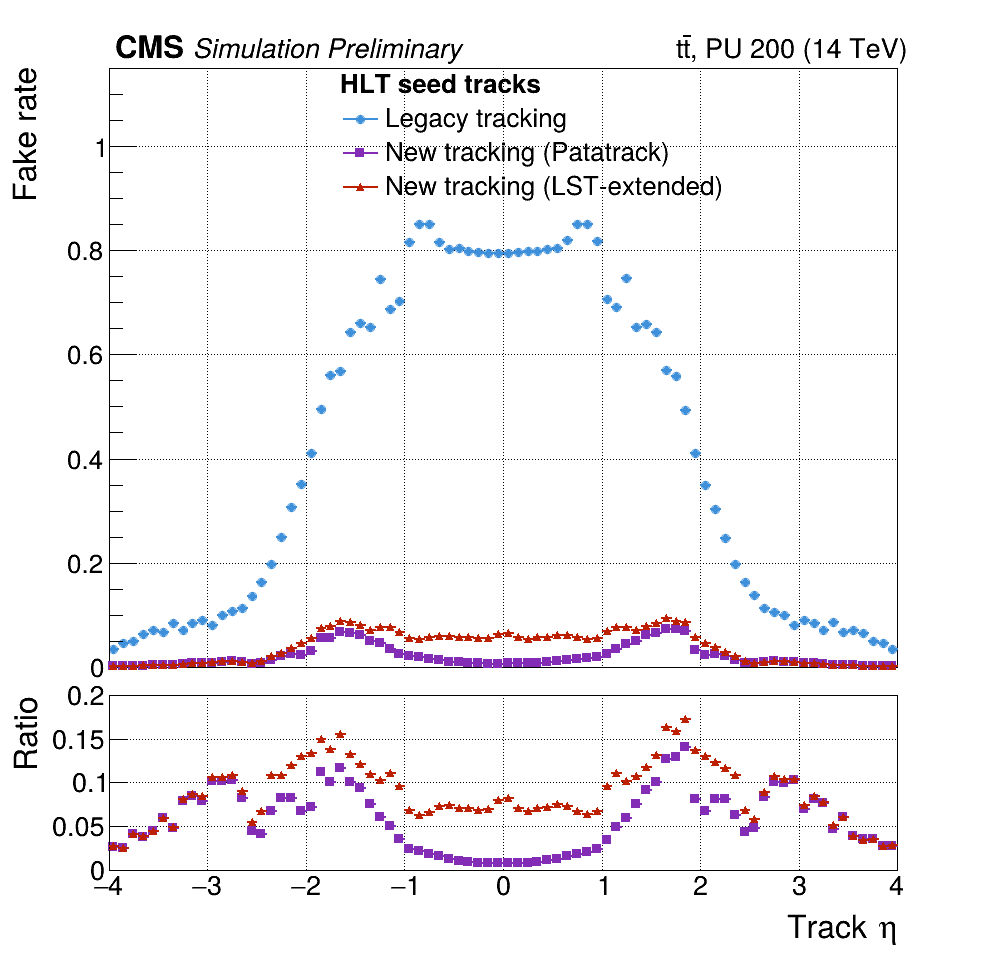}
  \caption{Efficiency of HLT seed tracks as a function of the simulated track production
  radius (left) and fake rate as a function of the track pseudorapidity (right), for legacy
  tracking, the new tracking with Patatrack seeding, and the new tracking with
  LST-extended seeding. From Ref.~\cite{dp026}.}
  \label{fig:seeds}
\end{figure}

\begin{figure}[htbp]
  \centering
  \includegraphics[width=0.32\textwidth]{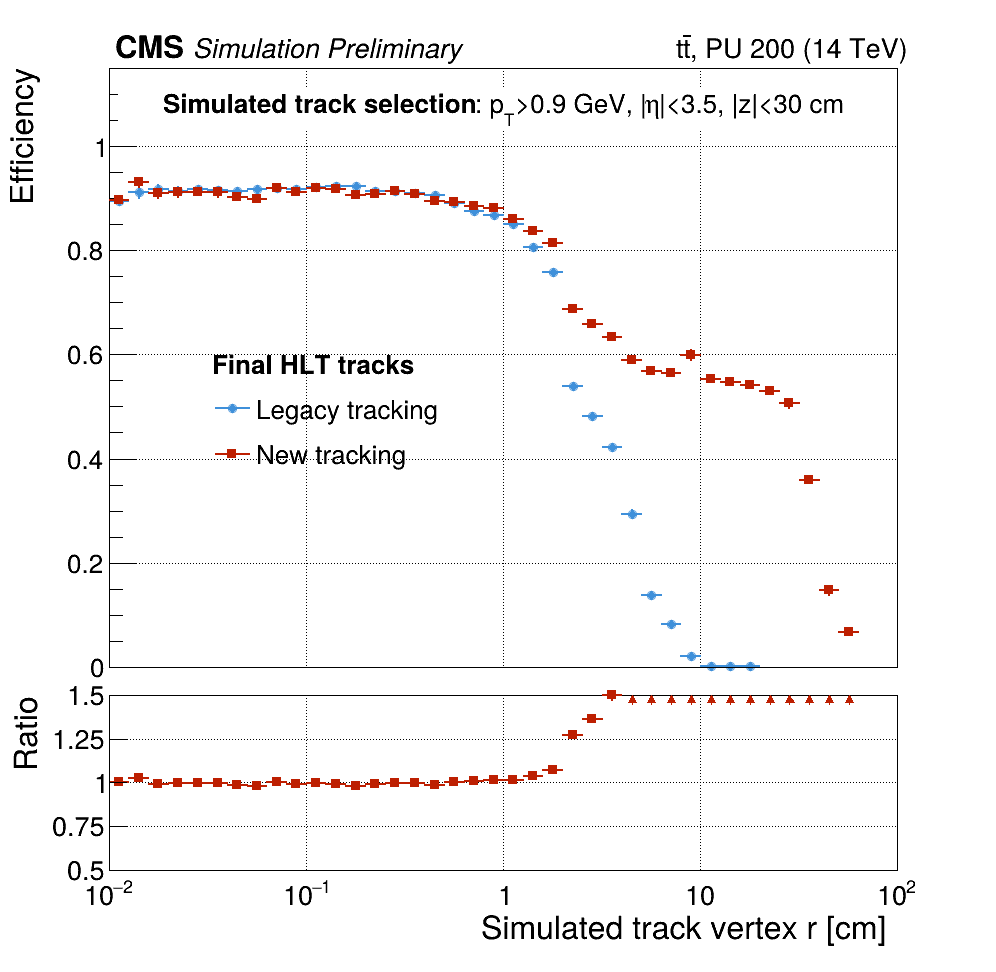}
  \includegraphics[width=0.32\textwidth]{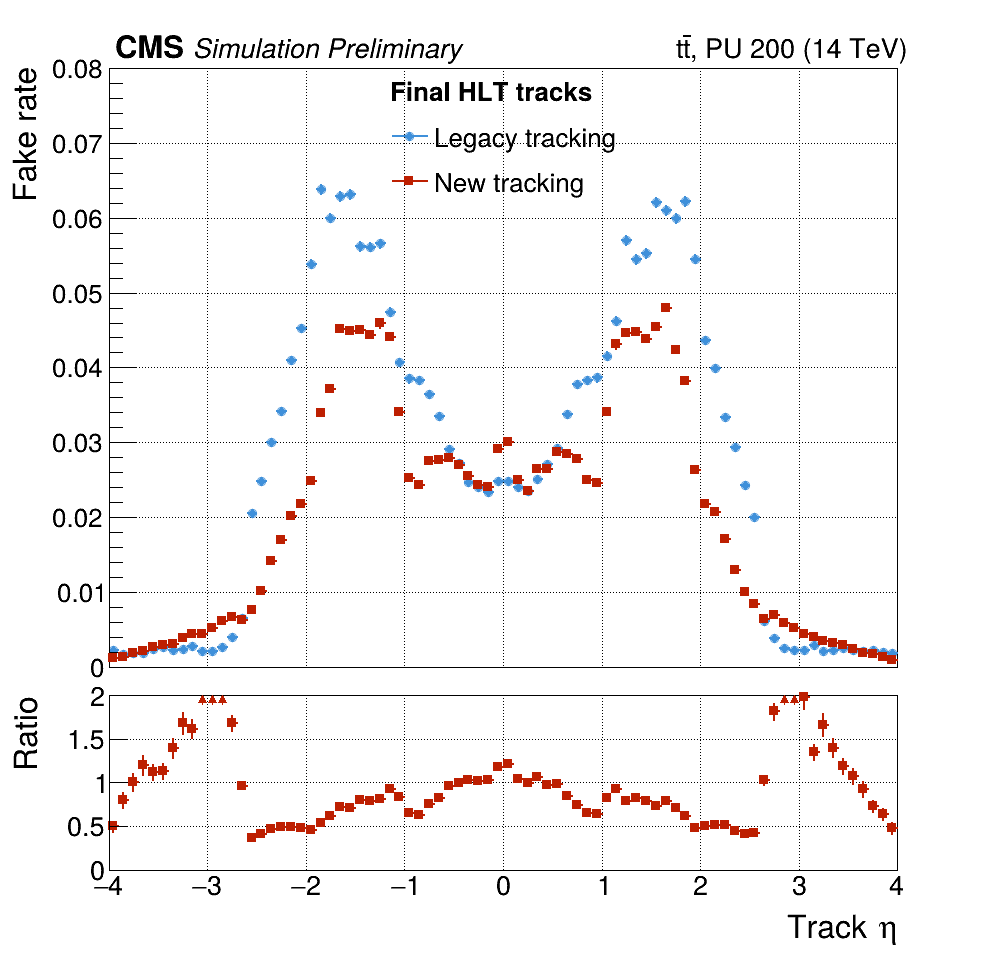}
  \includegraphics[width=0.32\textwidth]{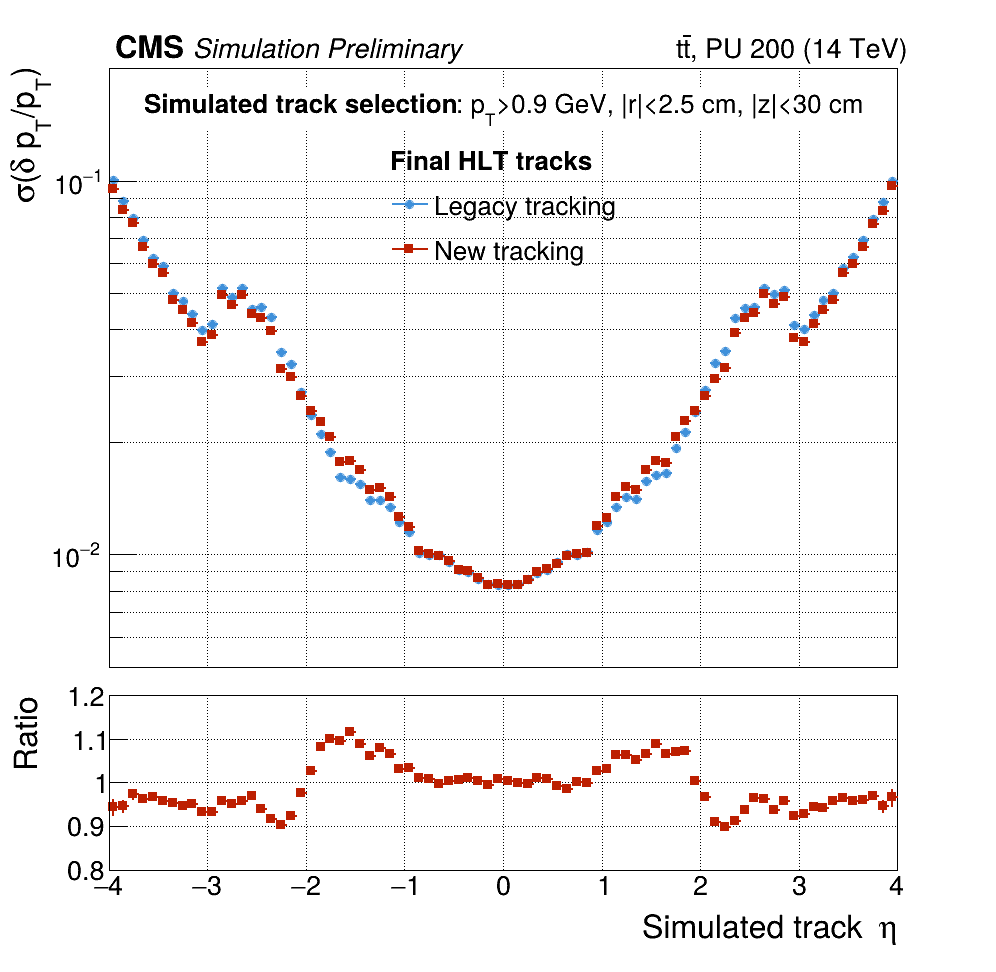}
  \caption{Final HLT tracks in the legacy tracking and in the new sequence: efficiency as a function of
  the simulated track production radius (left), fake rate as a function of the track
  pseudorapidity (middle), and $p_{\mathrm{T}}$ resolution as a function of the
  simulated track pseudorapidity (right). From Ref.~\cite{dp026}.}
  \label{fig:final}
\end{figure}

Figure~\ref{fig:seeds} compares seed tracks in the legacy and new HLT sequences.
Replacing the legacy IT-only seeding with Patatrack suppresses fakes
and duplicates by roughly an order of magnitude, at the cost of a slight
inefficiency. Adding LST extends the efficiency at large production radii 
at the cost of a small increase in fake rate with respect to Patatrack alone.
The same picture holds for the final tracks (Fig.~\ref{fig:final}):
mkFit preserves the gains of the new seeding while further reducing the fake
rate, achieving a $p_{\mathrm{T}}$ resolution competitive with the legacy sequence 
over the full pseudorapidity range.

The computational gains are also significant~\cite{dp026}.
When running with (without) GPU offloading,
the new sequence reduces the time spent in tracking by 64\% (35\%) 
and the overall HLT reconstruction time per event by 33\% (9\%).

Within the new baseline, track fitting has become as expensive as track building,
which makes it the natural next target:
extending mkFit to the fitting step~\cite{dp031} reduces the time spent
in fitting and format conversion by 71\%,
and shortens the full tracking sequence by 32\% (14\%)
when the rest of the sequence runs with (without) GPU offloading.

\section{Machine learning in Phase-2 tracking}

Machine Learning is already an established ingredient of CMS tracking.
In the new Phase-2 HLT sequence,
DNNs are used to pre-select seed tracks for Patatrack and LST,
and soon to select the final tracks, with inference on either CPUs or GPUs.

\begin{figure}[htbp]
  \centering
  \includegraphics[width=0.39\textwidth]{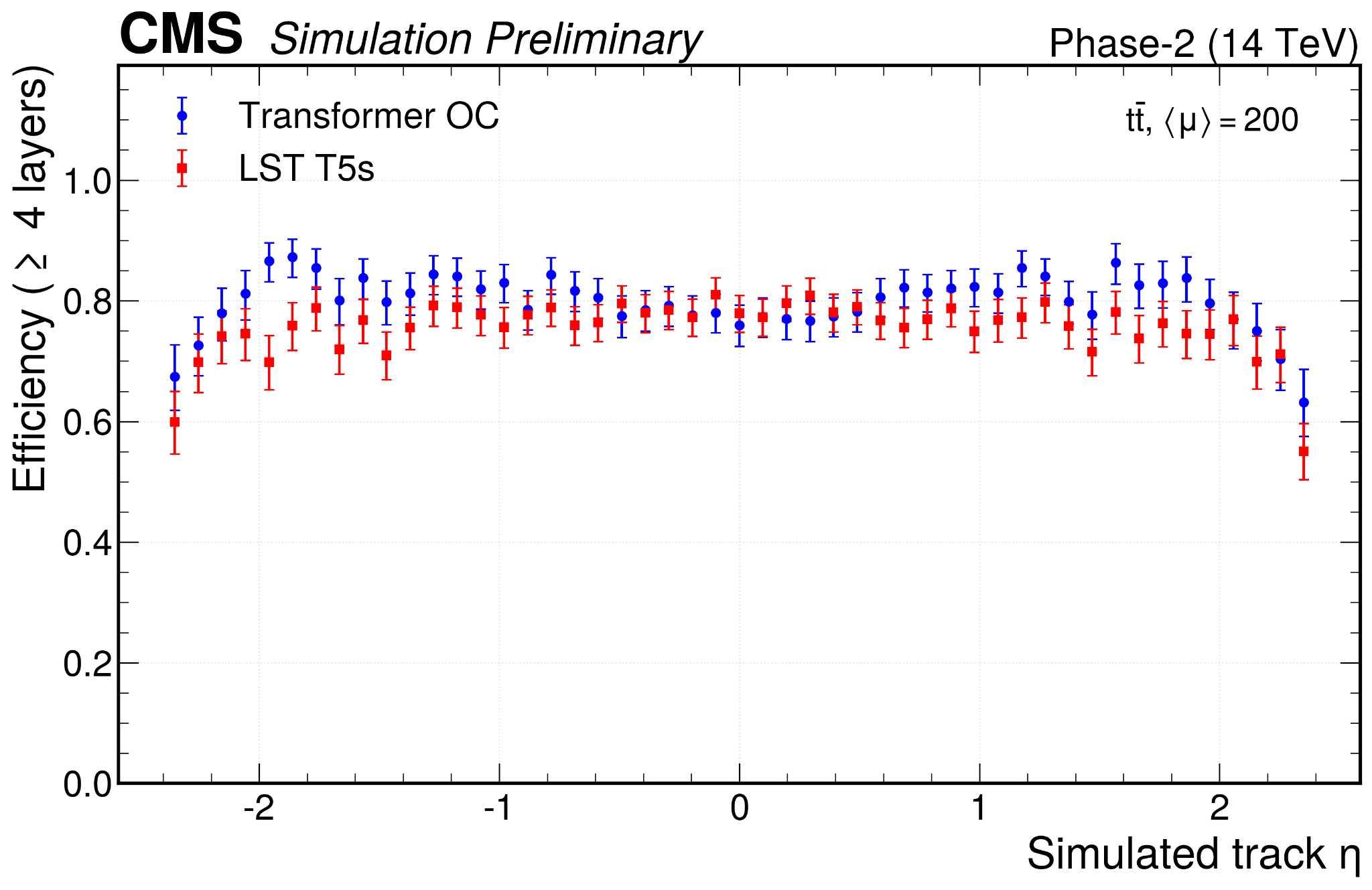}
  \includegraphics[width=0.39\textwidth]{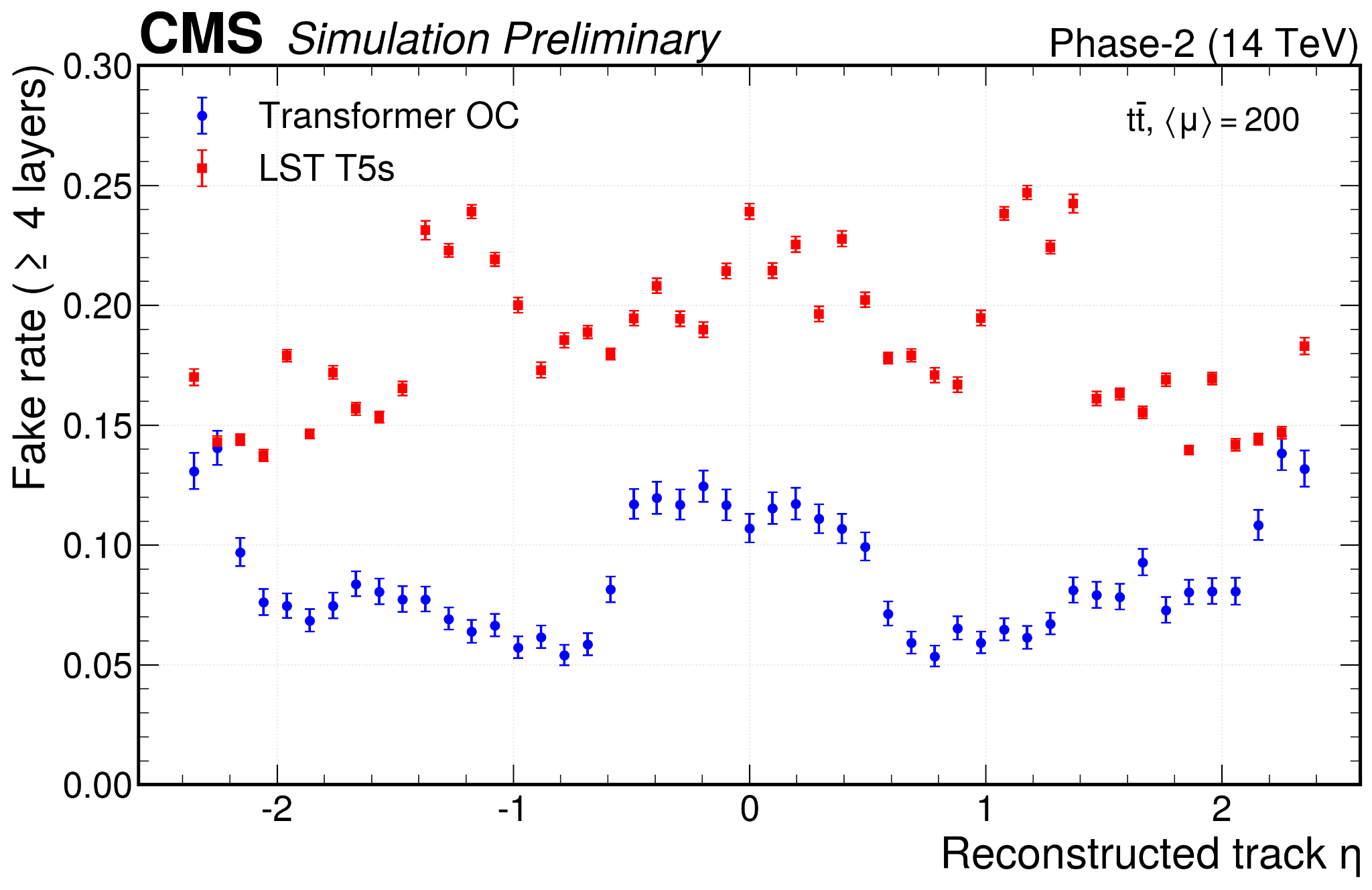}
  \caption{Efficiency for tracks crossing at least four layers as a function of the
  simulated track pseudorapidity (left) and fake rate as a function of the reconstructed
  track pseudorapidity (right), for the transformer-based
  reconstruction and for LST quintuplets (T5s). From Ref.~\cite{dp030}.}
  \label{fig:oc}
\end{figure}

Beyond such applications,
a more fundamental ML tracking approach is also being pursued,
using LST as its foundation~\cite{dp030}.
A transformer encoder produces an embedding for every LST triplet (T3)
within an object condensation loss framework: T3s belonging to the same particle are pulled together in the
embedding space while the others, including fake T3s, are pushed apart, 
each T3 being assigned a condensation likelihood alongside its coordinates.
The embeddings are then clustered into track candidates with DBSCAN~\cite{dbscan}.
These first results in CMS (Fig.~\ref{fig:oc}) are encouraging: the transformer-based reconstruction
matches or exceeds the efficiency of LST quintuplets (T5s)
while roughly halving the fake rate over the full pseudorapidity range.

In a similar vein, a slot-attention graph neural network
was recently used for vertex clustering
in a four-dimensional vertexing algorithm
that exploits the timing information of the new MIP Timing Detector~\cite{mtd},
with similarly encouraging results~\cite{dp042}.

\section{Summary and outlook}

Coping with up to 200 pileup interactions within a finite computing budget
requires improving physics and computational performance at the same time.
The Phase-2 CMS tracking does so by unifying GPU-friendly algorithms (Patatrack, LST), vectorized CPU algorithms
(mkFit) and machine learning (ML).
It retains essentially the full physics performance of the legacy sequence in the original phase space,
significantly extends the accessible phase space to displaced tracks, and significantly
reduces the reconstruction time per event.
Development continues on all fronts, to further improve both physics performance and timing
for track and vertex reconstruction at the HL-LHC.


\acknowledgments{This work was supported by the National Science Foundation under Cooperative Agreements OAC-1836650 and PHY-2323298.}



\begin{thebibliography}{99}
%
\bibitem{tktdr} CMS Collaboration, \emph{The Phase-2 Upgrade of the CMS
  Tracker}, \href{https://cds.cern.ch/record/2272264}{CERN-LHCC-2017-009,
  CMS-TDR-014} (2017).
%
\bibitem{dp026} CMS Collaboration, \emph{A heterogeneous \& vectorized sequence
  for the Phase-2 HLT tracking reconstruction},
  \href{https://cds.cern.ch/record/2961608}{CMS-DP-2026/026} (2026).
%
\bibitem{patatrack} A.~Bocci, V.~Innocente, M.~Kortelainen, F.~Pantaleo and
  M.~Rovere, \emph{Heterogeneous reconstruction of tracks and primary vertices
  with the CMS pixel tracker}, \href{https://doi.org/10.3389/fdata.2020.601728}{Front. Big Data \textbf{3}
  (2020) 601728}.
%
\bibitem{dp2022hlt} CMS Collaboration, \emph{Performance of Run-3 HLT track
  reconstruction}, \href{https://cds.cern.ch/record/2814111}{CMS-DP-2022/014}
  (2022).
%
\bibitem{dp2025pixel} CMS Collaboration, \emph{Performance of CMS pixel
  tracking at the High-Level Trigger for Phase-2},
  \href{https://cds.cern.ch/record/2948318}{CMS-DP-2025/076} (2025).
%
\bibitem{dp114} CMS Collaboration, \emph{Accelerated tracking for a
  high-throughput Phase-2 NGT HLT Scouting stream: computing and physics
  performance}, \href{https://cds.cern.ch/record/2967245}{CMS-DP-2026/114}
  (2026).
%
\bibitem{dp029} CMS Collaboration, \emph{Extending the reconstruction of
  Phase-2 displaced tracks using Line Segment Tracking},
  \href{https://cds.cern.ch/record/2961685}{CMS-DP-2026/029} (2026).
%
\bibitem{dp2025lstml} CMS Collaboration, \emph{Extended ML selections and track
  embeddings for duplicate removal in Line Segment Tracking (LST)},
  \href{https://cds.cern.ch/record/2941435}{CMS-DP-2025/048} (2025).
%
\bibitem{mkfit} S.~Lantz et al., \emph{Speeding up particle track
  reconstruction using a parallel Kalman filter algorithm}, \href{https://doi.org/10.1088/1748-0221/15/09/P09030}{JINST \textbf{15} (2020) P09030}.
%
\bibitem{dp2022mkfit} CMS Collaboration, \emph{Performance of Run-3 track
  reconstruction with the mkFit algorithm},
  \href{https://cds.cern.ch/record/2814000}{CMS-DP-2022/018} (2022).
%
\bibitem{dp031} CMS Collaboration, \emph{Application of mkFit to track fitting
  for CMS Phase-2 tracking at the HLT},
  \href{https://cds.cern.ch/record/2961687}{CMS-DP-2026/031} (2026).
%
\bibitem{dp030} CMS Collaboration, \emph{Using Line Segment Tracking to enable
  ML-based track reconstruction in Phase-2 of the CMS experiment},
  \href{https://cds.cern.ch/record/2961686}{CMS-DP-2026/030} (2026).
%
\bibitem{dbscan} M.~Ester, H.-P.~Kriegel, J.~Sander and X.~Xu, \emph{A
  density-based algorithm for discovering clusters in large spatial databases
  with noise}, in \href{https://cdn.aaai.org/KDD/1996/KDD96-037.pdf}{Proceedings of the Second International Conference on
  Knowledge Discovery and Data Mining (KDD'96)}, AAAI Press (1996), p.~226.
%
\bibitem{mtd} CMS Collaboration, \emph{A MIP Timing Detector for the CMS
  Phase-2 Upgrade}, \href{https://cds.cern.ch/record/2667167}{CERN-LHCC-2019-003,
  CMS-TDR-020} (2019).
%
\bibitem{dp042} CMS Collaboration, \emph{GNN based primary vertex
  reconstruction with slot attention},
  \href{https://cds.cern.ch/record/2961848}{CMS-DP-2026/042} (2026).
%
\end{thebibliography}
\end{document}